\documentclass[11pt]{article}  

\DeclareSymbolFont{AMSb}{U}{msb}{m}{n}
\DeclareMathSymbol{\N}{\mathbin}{AMSb}{"4E}
\DeclareMathSymbol{\Z}{\mathbin}{AMSb}{"5A}
\DeclareMathSymbol{\R}{\mathbin}{AMSb}{"52}
\DeclareMathSymbol{\Q}{\mathbin}{AMSb}{"51}
\DeclareMathSymbol{\I}{\mathbin}{AMSb}{"49}
\DeclareMathSymbol{\C}{\mathbin}{AMSb}{"43}

\title{Causal sets and the deep structure of spacetime}
\author{Fay Dowker
\\
Blackett Laboratory, Imperial College, London, SW7 2AZ, UK.
}
\begin{document}
\maketitle
\begin{abstract}
The causal set approach to quantum gravity embodies the concepts of 
causality and discreteness. This article explores some foundational
and conceptual issues within causal set theory.  

\end{abstract}

\section{Introduction}     

The problem of quantum gravity in its narrow sense is the problem
of finding a theory that incorporates both Quantum
Mechanics and General Relativity. A broader vision is that a theory 
of quantum gravity would restore to physics a unified framework, a
framework in which there is no fundamental division in principle
between observer and observed nor between matter and spacetime.
One reading of Einstein's writings on Quantum Mechanics and
Unification is that he viewed success in the broad quest as a
prerequisite to a solution of the narrow problem.

Every approach to quantum gravity in this broad conception must
embody answers to two fundamental questions: ``What is quantum
mechanics?'' and ``What is the deep structure of spacetime?'' 
This article will touch on the former question 
and focus on the latter and the answer to it
provided by  the approach known as causal set theory  
which marries the two concepts of
discreteness (or atomicity) and causality.

The view that causality is a more fundamental organising
principle, even than space and time, is an ancient tradition
of thought.   
Corresponding to this view, momentary events have a
better claim to be basic than objects extended in time like particles,
the latter being understood as persistent patterns of events
rather than enduring ``substances''. 
Within Relativity, the recognition that
almost all of the geometrical properties of Minkowski space could
be reduced to order theoretic relationships among point events
came very early.\cite{Robb:1936}

The concept of atomicity also has a long history as do
philosophical challenges to the antithetical notion of a
physical continuum. Of course Quantum Mechanics 
itself is named for the discreteness of atomic and subatomic phenomena. In
more recent times, the increasing importance of computing with its
discrete algorithms and digital processing has had a pervasive
influence on intellectual culture. Whatever the roots of the
concept, it is certainly now the case that many workers 
believe that a fundamentally discrete structure to reality is
required to reconcile spacetime with the quantum. Einstein himself
adumbrated this view and it is impossible to resist
the temptation of quoting him here:

``But you have correctly grasped the drawback that the continuum
brings. If the molecular view of matter is the correct
(appropriate) one, {\it i.e.}, if a part of the universe is to be
represented by a finite number of moving points, then the
continuum of the present theory contains too great a manifold of
possibilities. I also believe that this too great is responsible
for the fact that our present means of description miscarry with
the quantum theory. The problem seems to me how one can formulate
statements about a discontinuum without calling upon a continuum
(space-time) as an aid; the latter should be banned from the
theory as a supplementary construction not justified by the
essence of the problem, which corresponds to nothing ``real''. But
we still lack the mathematical structure unfortunately. How much
have I already plagued myself in this way!''

\noindent A. Einstein in a letter to Walter D\"allenbach, November 1916, 
translated and cited by John Stachel.\cite{Stachel:1986}

Causal set 
theory\cite{tHooft:1979,Myrheim:1978,Bombelli:1987aa,Sorkin:1990bh,Sorkin:1990bj} 
arises by combining discreteness and causality 
to create a substance that can be the basis
of a theory of quantum gravity. Spacetime is thereby replaced by a
vast assembly of discrete ``elements'' organised by means of
``relations'' between them into a ``partially ordered set'' or
``poset'' for short. None of the continuum attributes
of spacetime, neither metric, topology nor differentiable 
structure, are retained, but emerge it is hoped as 
approximate concepts at macroscopic scales. 

Amongst current approaches, causal set theory can claim to lie at the
extreme end of the granularity scale:
it is maximally discrete. No continuum concept is required to
specify the underlying reality which is purely
combinatorial. The elements have no internal structure; they are
the fundamental units of reality. All one can do is count: count
elements, count relations, count certain patterns of elements and
relations. The slogan might be coined, ``Real numbers are not real
in a causal set.''

The hypothesis that the deep structure of spacetime is a discrete
poset characterises causal set theory at the {\it kinematical}
level; that is, it is a proposition about what substance is the
subject of the theory. However, kinematics needs to be completed
by {\it dynamics}, or rules about how the substance behaves, if
one is to have a complete theory.
In  this article I 
will explore some foundational issues within
these two categories in a non-technical way. I have not 
attempted to give anything approaching a review of 
causal set theory. Reference 
\cite{Sorkin:2003bx}, by Rafael Sorkin who has done more
than anyone to further the causal set approach, gives more details of  
current developments.

\section{Kinematics}

\subsection{The causal set}

Mathematically, a causal set is defined to be a locally finite partially
ordered set, or in other words a set $C$ together with a relation $\prec$, 
called ``precedes'', which satisfy the following axioms:

\noindent (1) if $x \prec y$ and $y \prec z$ then $x \prec z$, 
$\forall x,y,z \in C$ (transitivity);

\noindent (2) if $x \prec y$ and $y\prec x$ then 
$x = y$ $\forall x,y \in C$ (non-circularity);

\noindent (3) for any pair of fixed elements $x$ and $z$ of $C$, the set 
$\{y | x \prec y \prec z \}$ of elements lying between $x$ and $z$ is finite. 

Of these axioms, the first two say that $C$ is a partially ordered
set or poset and the third expresses local finiteness. 
The idea is that in the deep quantum regime of very small distances, 
spacetime is no longer described by a tensor field, the metric, 
on a differentiable manifold, but by a causal set. The discrete elements 
of the causal set  are related to each other only by the partial 
ordering that corresponds to a microscopic notion of before and after, 
and the continuum notions of length and time arise only as approximations 
at large scales. 

The richness of the structure of partial orders is reflected in the
many different sets of terminologies used by mathematicians and
physicists who study them. One of the most useful and suggestive for
the purposes of quantum gravity is the ``genealogical'' jargon 
whereby one thinks of a causal set as a family tree. An element 
$x$ is an {\it ancestor} of an element $y$ if $x \prec y$, 
and $y$ is then a {\it descendant} of $x$. 

To arrive at this structure as the kinematical basis for
quantum gravity, one can start by conjecturing discreteness 
and see how this leads to the addition of causal structure, or vice 
versa. The scientific arguments for the two interweave each
other and it's an artificial choice. Indeed, 
it is the fact that the two concepts ``complete'' each other in the
context of a proposed substructure for spacetime that is one of
the strongest motivations for the causal set programme. 
For the purposes of the current exposition let's choose to begin 
by postulating a discrete spacetime. 

\subsection{An analogy: discrete matter}

In quantum gravity, we know
what the continuum description of spacetime is at macroscopic scales
-- it is a Lorentzian manifold --  
 and are trying to discover the
discrete substratum. It is useful to imagine
ourselves in the analogous situation 
for a simple physical system. Consider a quantity of 
a material substance in a box in flat space for which 
we have a continuum description -- a mass density, say -- but which we 
suspect is fundamentally discrete. The question is, ``What could
the discrete state be that gives rise to this continuum approximation?''  
and a good first response is to try to {\it discretise} the 
continuum. We then try to give the discrete object so created
``a life of its own'' as an entity independent of the continuum 
from which it was derived. We then ask whether we can believe in  
this discrete object as fundamental and 
whether and how we can recover from it a continuum approximation 
which may not be exactly the original continuum from which we 
started but which must be ``close'' to it. 

In the case of the material substance, let us postulate that it is 
made of identical atoms. The varying mass density 
then can only be due to differing number densities
of atoms in space. We discretise by 
somehow distributing the atoms in the box 
so that the number of atoms in each sufficiently 
large region is approximately proportional 
to the mass density integrated over that region. We may or may 
not have reasons to suspect what the atomic mass actually is, 
but in any case it must be small enough so that the spacing 
between the atoms is much smaller than the scales on which the 
density varies. 

Each method of discretising will produce distributions of atoms that have 
different properties and which type is more favourable as a fundamental 
state depends on which features of the continuum theory we wish 
to preserve and how fruitful each turns out to be when 
we come to propose dynamics for the discrete structure
itself.  Suppose for example that the 
continuum theory of the material is invariant under Euclidean transformations
(rotations and translations) at least locally in small enough 
regions over which the mass density is approximately constant
and ignoring edge effects. There are ways to discretise which do not 
respect the invariance under Euclidean transformations. For example, 
we can divide the box into cubes small enough so the density is 
approximately constant in each. In each cube, place an atom at every vertex
of a Cartesian lattice with lattice spacing chosen inversely proportional
to the mass density in the cube. We can produce in this way an atomic 
state from which we can recover approximately the correct 
continuum mass density but  
it is not invariant under the Euclidean group -- there are 
preferred directions -- 
and if a fully fledged fundamental discrete theory is based on such 
lattice-like atomic states, this will show up in deviations from 
exact Euclidean invariance in the continuum approximation to this 
full underlying theory. 
 
If a discretisation that respects the invariance is desired, 
there must be no preferred directions and
this can be achieved by taking a {\it random} distribution of atoms, 
in other words atomic positions chosen according to a Poisson distribution 
so that the expected number of atoms 
in a given region is the mass in that region (in atomic mass units).  
This will produce, with high probability, an atomic configuration that 
does not distinguish any direction.  

Having placed the atoms down, we kick away the prop
of the continuum description
and ask if the distribution of atoms itself could be the underlying 
reality; in particular how do we start with it and 
recover, approximately, a continuum description? 
To answer this question, we can use the discretisation 
in reverse: a continuum mass density is a good approximation to 
an atomic state if that atomic state could be a discretisation of 
the mass density. In the case of the atomic states arising from random 
discretisations, this is modified to: a mass density 
is a good approximation if the atomic state could have arisen with 
relatively high probability from amongst the possible discretisations.
Then it must be checked that  
if two continua are good approximations to the same 
atomic state, they must be close to each other -- this 
is crucial if this relationship between 
discrete states and continuum states is to be consistent.

Finally, we propose the atomic state as the underlying reality, reinterpret
{\it mass} as a measure of the {\it number} of atoms in a region,
 ask whether there are atomic states which have no 
continuum approximation and what their meaning is, start working on a
dynamics for the atoms {\it etc.} 

\subsection{Discrete spacetime}

These steps are straightforward in this simple case, and
 can be taken in analogous fashion in the discretisation of a
continuum spacetime given by a Lorentzian metric tensor on a differentiable 
manifold. Let us tread the same path. 

What plays the role of the mass density? A good candidate 
for the measure of the sheer quantity of spacetime in 
a region is the {\it volume} 
of that region. It is calculated by integrating over the 
region the volume density given by the square root of minus 
the determinant of the spacetime metric, $\int_{region} \sqrt{-g}\, d^4x$ 
and is a covariant quantity: it is independent 
of the coordinates used to calculate it. 

In the case of quantum gravity we have independent evidence, from 
the entropy of black holes for example,\cite{Sorkin:1997gi} that the 
scale of the discreteness is of the order of the Planck scale formed
from the three fundamental constants, $G$, $\hbar$ and $c$. If the Planck 
length is defined to be $l_p = \sqrt{\kappa \hbar}$ where $\kappa = 
8 \pi G$ and we have set $c = 1$, 
then the fundamental unit of volume 
will be $V_f \equiv \nu l_p^4$ 
where $\nu$ is a number 
of order one and yet to be determined.
In order to discretise a spacetime, we distribute 
the atoms of spacetime, which we will call simply ``elements'',
in the spacetime in such a way that the number of them 
in any sufficiently large region is approximately  
the volume of that region in fundamental volume units. 

Analogous to the question of Euclidean invariance in the material example,
we must ask: do we want to preserve Lorentz invariance or not?
We might think of laying down a grid of coordinates on spacetime
marked off at $\nu^{1/4} l_p$ intervals and placing an element at every 
grid vertex. The problem that immediately arises is that 
it is not a covariant prescription and this manifests itself 
dramatically 
in frames highly boosted with respect to the frame
defined by the coordinates, where the distribution of elements 
will not be uniform but will contain large void regions 
in which there are no elements at all (see {\it e.g.} reference
\cite{Dowker:2003hb} for a picture). 
Such a coordinate-dependent discretisation will therefore violate
Lorentz invariance. 
This 
is a much more serious matter than the  breaking of Euclidean invariance
for a coordinate-based discretisation of the material. The breaking  
of Lorentz invariance manifests itself by a failure of the 
distribution of elements to be uniform {\it at all} in highly boosted frames.
If such a discrete entity were to be proposed as
fundamental, it would have to be concluded that in
highly boosted frames there can be no 
manifold description at all. 
 
There is as yet no evidence that Lorentz invariance is
violated so let us try to maintain it in quantum gravity. In seeking a 
Lorentz invariant discretisation process 
the crucial insight is, again, that the discretisation should be
random.\cite{Bombelli:1987aa} It is performed via a 
process of ``sprinkling'' which is the 
Poisson process of choosing countable subsets 
of the spacetime for which the expected
number of points chosen from any given region of spacetime
is equal to its volume in fundamental units.  
That this process is exactly Lorentz invariant in Minkowski spacetime
is a consequence of the fact that the Minkowskian volume element is 
equal to the Euclidean volume element on $\R^n$ and the 
fact that the Poisson process in $\R^n$ is invariant under any 
volume preserving map (see {\it e.g.}
reference \cite{Stoyan:1995}).

The sprinkling process results in a uniform 
distribution of elements in the spacetime
but this set of discrete elements cannot, by itself, provide a
possible fundamental description of spacetime. Here, our analogy
with a simple mass density in a box breaks down. When, having constructed a
distribution of atoms in the box, we take away the
continuum mass density, like whisking away the tablecloth from 
under the crockery, the atoms retain their
positions in space because Euclidean space is an assumed
background -- the table -- for the whole setup. But in the case of 
quantum gravity, the sprinkled elements are meant to {\it be}
the spacetime and if we whisk spacetime away from
the elements we have sprinkled into it
-- removing the table not just the tablecloth -- the elements 
just collapse into a heap of unstructured dust. 
The sprinkled elements must be endowed with some extra structure.

We already know that by counting
elements we can recover volume information in the continuum 
approximation. Powerful theorems in causal 
analysis\cite{Hawking:1976fe,Malament:1977,Levichev:1987}
show that what is needed in the continuum to complete
volume information to give the full spacetime geometry is the 
{\it causal structure} of a spacetime.\footnote{The theorems apply to 
spacetimes that satisfy a certain global causality condition --
{\it past and future distinguishability} -- which means that 
distinct points have distinct causal pasts and futures and which 
we will assume for every continuum spacetime referred to here.
We can say that the theorems imply that causal set theory
predicts spacetime must satisfy this condition because 
only such spacetimes will be able to approximate a causal set.}

The causal structure of a spacetime is the totality of the information about 
which events can causally influence which other events. For each 
point $p$ of the spacetime, we define the set $J^-(p)$ ($J^+(p)$)
the causal past (future) of $p$, to be the set of points, $q$, in the 
spacetime for which there is a future (past) directed causal curve
-- a curve with an everywhere non-spacelike, future (past) pointing 
tangent vector -- from $q$ to $p$. The collection of all these causal past and 
future sets is the causal structure of the spacetime. This is often
colloquially called 
the ``light cone structure'' because the boundaries of these sets
from a point $p$ are the past and future lightcones from $p$ and 
the sets themselves are the lightcones and their interiors.  

Let us therefore, in our discretisation procedure,
endow the elements sprinkled into the spacetime with 
the order given by the spacetime causal structure: 
elements $e_i \prec e_j$ if they are sprinkled at points 
$p_i$ and $p_j$ respectively in the 
continuum such that $p_i \in J^-(p_j)$. The set of sprinkled
elements with this induced order is a 
causal set satisfying the axioms given above.

Now we give the causal set independence from the spacetime.
To recover from it approximately the continuum we discretised, 
we follow the guidance of the example of the substance in a box: 
a spacetime $M$ is a good approximation to a causal set 
$C$ if $C$ could have arisen from $M$ 
by the  discretisation process we have described (sprinkling and endowing with 
induced order) with relatively high probability. 

As mentioned in the material-in-a-box example, consistency 
requires that if two continua are good approximations to the 
same discretum, they should be close to each other. This is 
a central conjecture, the {\it Hauptvermutung}, of causal set theory
and it is surprisingly hard even to formulate it precisely 
due to the difficulty of defining a 
notion of distance on the space of Lorentzian manifolds. 
We use the intuitive idea constantly -- how else would it 
make sense to talk of one spacetime being a small perturbation of
another -- but only recently has 
progress been made in this direction.\cite{Bombelli:2000wu,Noldus:2003si}
This progress has been inspired by causal sets, 
in particular by utilising and comparing the 
different probability distributions on the space of
causal sets given by the sprinkling processes into 
different Lorentzian manifolds. If it is the case that 
using sprinklings is the only way covariantly to say what we mean by two 
Lorentzian spacetimes being close,  it would be further 
evidence for causal sets as the deep structure.

\subsection{Reassessed in a quantal light}

It may be argued that the above steps 
leading to the proposal of causal sets for quantum gravity kinematics 
have been taken under the assumption that the continuum 
is an approximation to one single discrete 
spacetime whereas Quantum Mechanics would suggest that a continuum spacetime
is better characterised as corresponding to
a coarse grained set of many discreta.   
This point has validity and in addressing it, we are drawn into the 
realm of dynamics and the question of what form
a quantum dynamics for causal sets might take, the subject of the 
next section. 
Certainly, the statement that the number and ordering of causal 
set elements correspond 
to continuum volume and causal structure, respectively, 
will have to be judged and interpreted in the light of the 
full dynamical theory. 
In the meantime, however, Quantum Mechanics need 
make us no more squeamish
about the statement, ``Spacetime is a causal set'', than
about the statement, ``Things are made of atoms.'' 

Even if it turns out that only a whole bunch of discreta 
can properly be said to correspond
to a continuum spacetime, we can still make the claim that 
discrete data can give rise, 
in a Lorentz invariant manner, to a continuum spacetime 
if they are organised as a 
causal set. In this case, we would say that the data  common 
to, or shared by, 
each member of the bunch of discreta -- a coarse graining of them all -- 
is a causet. The question would then be, what discretum
can be coarse grained to give a causet?

The answer is, another ``finer'' causet! Indeed causets admit a notion of
coarse graining that is consistent with the inverse procedures of
discretisation and continuum approximation given above because it is
itself a random process. To perform a $\frac{1}{q}: 1$ coarse 
graining of a causet, go through it 
and throw away each element with fixed probability $p = 1-q$. A causet $a$
is a coarse graining of a causet $b$ if $a$ could have resulted
with relatively high probability from the process of coarse graining 
$b$.

On this view, a spacetime $M$ could correspond dynamically to 
a set of ``microscopic'' states which are causets with 
no continuum approximation at all, but which have a common coarse
graining to which $M$ {\it is} a good approximation.

It should be mentioned that this process of
coarse graining allows the notion of scale dependent topology 
to be realised in causal set theory. Many quantum gravity 
workers have the intuitive picture that at scales close to the 
Planck scale, the topology of spacetime will be 
complicated by many wormholes and other nontrivial topological 
excitations, but at every day  scales the topology is
trivial. As attractive as this idea is, I know of no way in the
continuum to make it concrete. If spacetime is a 
causet, however, coarse graining it at different scales ({\it i.e.}
with different deletion probabilities $p$) gives 
rise to different causets which may have continuum approximations
with different topologies, including different dimensions.

\section{Dynamics}

If we imagine a bag containing all the causets with $N$ elements, 
where $N$ is some immensely large number, then drawing out of the bag
a causet uniformly at random will result, with probability approaching 1 as
$N$ tends to infinity, in a causet with a very specific and rather
surprising structure.  
This structure is of three ``levels'': the first level is of elements 
with no ancestors and has roughly $N/4$ members, the second is of 
elements with ancestors in level 1 and has roughly $N/2$ members and 
the remaining elements are in level 3 with ancestors in 
level 2.\cite{Kleitman:1975}

These 3-level causets have no continuum approximation -- they are 
universes that ``last'' only a couple of Planck times.
If the fundamental 
reality is a causal set then we have to explain why the causet that 
actually occurs is a manifold-like one and not one of
the, vastly more numerous, 3-level causets. 
This is the causal set version of 
a problem that is common to all discrete approaches
to quantum gravity: the sample space of discreta
is always dominated in sheer numbers 
by the non-manifold-like ones and a uniform distribution 
over the sample space will render these bad ones overwhelmingly 
more likely. 
We need a dynamics to cancel out this ``entropic''
effect and to produce a measure over the sample space 
that is peaked on the manifold-like entities.   

As mentioned in the introduction, the broad goal of quantum
gravity includes the unification of observer with observed
(or rather the elimination of the observer as
fundamental). Thus, finding 
a dynamics for quantum gravity, involves a resolution of the vexed problem 
of the ``Interpretation of Quantum Mechanics.'' 
There is no consensus on how
this is to be achieved and this difficulty could be seen as a severe 
obstacle in the quest for quantum gravity. 
Turning the problems around, however, the requirements of quantum gravity, 
for example general covariance, might be taken as a guide 
for how to approach Quantum Mechanics in general.
Indeed, quantum gravity points to an approach 
to Quantum Mechanics that is based on the histories of a system.

\subsection{A histories framework for quantum causal sets}

The reader may already be aware that the terms ``quantum state''
and ``Hilbert space'' {\it etc.} have been conspicuous by their absence in 
the present account. I have avoided them for reasons bound up 
with our broad goals in quantum gravity. The standard approach to 
Quantum Mechanics based on quantum states, {\it i.e.} elements 
of a Hilbert space, is tied up with the Copenhagen 
interpretation with its emphasis on ``observables'' and ``observers''.
The quantum state is a state ``at
a moment of time'' (in the Schr\"odinger picture) and to define it 
requires a foliation of spacetime by spatial hypersurfaces. Taking the 
principle of general covariance fully to heart suggests that  
our goals are better served by instead 
maintaining the fully spacetime character of reality
inherent in General Relativity, which 
points to the framework of the ``sum-over-histories'' for 
quantum gravity.\cite{Hawking:1978jz,Hartle:1992as,Sorkin:1997gi}

The bare bones structure of a histories quantum theory is 
a sample space, $\Omega$ of possible histories of the system in question 
--  for us, causal sets -- and a dynamics for the system expressed in terms 
of a ``quantum measure''  $\mu$, on $\Omega$. I will say more about 
the quantum measure below. For now, note that  
were $\mu$ a probability measure, this would be the 
familiar case of a classical stochastic theory and indeed a histories 
quantum theory can be thought of as a generalisation
of such a theory. That being so, we can prepare for 
our task of finding a quantum dynamics
by studying classical 
stochastic causets. 

\subsection{A classical warm up}

Just as a probability measure on the space of all paths on  
the integers -- a random walk --  can be given by 
all the ``transition 
probabilities'' from  each incomplete path, $\gamma$, to 
$\gamma$ plus an extra  
step, so a measure on the space of all possible past finite\footnote{Past
finite means that every element has finitely many ancestors.} 
causets can be specified by 
giving all the transition probabilities from 
each $n$-element causet, $c$, to all possible causets formed from 
$c$ by adding a single new element to it. The ancestors of the 
newly born element are chosen from the elements of 
$c$ according to the distribution of the transition 
probabilities. This is a process of 
stochastic ``sequential growth'' of a causet.\cite{Rideout:1999ub}

Without any restrictions, 
the number of such ``laws of growth'' is so huge that the class of them 
is not very useful to study. Imposing on the dynamics two physically motivated 
conditions severely narrows down the class, however. These conditions are
{\it discrete general covariance} and {\it Bell causality.} 
The first condition states that the probability of growing a 
particular finite partial causet does not depend on the order in which the 
elements are born. This is recognisably a ``label independence'' condition
and analogous to the independence of the gravitational action from 
a choice of coordinates. The second condition is 
the closest possible analogue, in this setting of 
causet growth, of the condition 
that gives rise to the Bell Inequalities in an ordinary stochastic theory 
in a background with fixed causal structure. It is meant to 
imply that a birth taking place in one region of the causet cannot 
be influenced by any birth in a region spacelike to the first. 

The solution of these two conditions is the class of ``Rideout-Sorkin''
models. Each model is specified by a sequence of non-negative real 
numbers, 
$t_0 = 1, t_1, t_2, \dots$ and  
the transition probabilities can be expressed in terms of these ``coupling
constants.'' The models are also known as ``generalised percolations''
since they generalise the dynamics called ``transitive percolation'' 
in which a newly born element independently 
chooses each of the already existing 
elements to be its ancestor  
with probability $p$. In generalised percolation a newly born element 
chooses a set of elements of cardinality $k$ to be its ancestors 
with relative probability $t_k$.\footnote{It would be more accurate to  
say ``proto-ancestors'' because the set has to be completed by adding 
all its own ancestors.}  
 
This family of models has proved to be a fruitful test bed for various
issues that will arise in the quantum theory, most notably that of
general covariance. 

\subsection{The problem of general covariance}

Part of the problem of general covariance is to identify 
exactly how the principle manifests itself in quantum gravity and 
how the general covariance of General Relativity arises from it. 
The answer will surely vary from approach to approach but some
overarching comments can be made. 
Part of what general covariance means is that physical quantities
and physical statements should be 
independent of 
arbitrary labels. 
It is needed as a principle when the dynamics of the system 
is not expressible (or not obviously expressible) directly in terms of the 
label invariant quantities, but can only be stated as a 
rule governing the labelled system. To atone for the sin of
working with these meaningless labels, the principle of general 
covariance is invoked and physical statements must be purged of these
unphysical markers. 

This is a difficult thing to do even in classical General 
Relativity. In quantum gravity, the issue is even  more fraught
because it is caught up in the quantum interpretational argy bargy. 
But the rough form of the problem can be understood in the 
following way. When the dynamics
is expressed in terms of labelled quantities, we can satisfy
general covariance by formally ``summing over all possible 
labellings'' to form covariant classes of things (histories, operators,
whatever). But in doing so, 
the physical meaning of those classes is lost, in other words  
the relation of these classes to  label-independent data 
is obscure. Work must then be done to recover this 
physical meaning. 
As an example, consider, in flat spacetime, the set of 
non-abelian gauge fields 
which are zero in a fixed spacetime region 
and all fields which 
are gauge equivalent to them.
This is a gauge invariant set of fields, but its 
physical meaning is obscure.
Uncovering a gauge invariant characterisation of 
elements of the set takes work.  

In causal set theory, progress has been made on this 
issue.\cite{Brightwell:2002yu,Brightwell:2002vw}
 A Rideout-Sorkin dynamics 
 naturally gives a measure on the 
space of {\it labelled} causets, $\tilde\Omega$ -- the labelling is the 
order in which the elements are born. This labelling is 
unphysical, the only physical structure possessed 
by a causet is its partial order. General covariance 
requires that only the covariant subsets of $\tilde\Omega$ have 
physical meaning where a covariant subset is one which if it 
contains a labelled causet, also contains all other labellings 
of it. Such subsets can be identified with subsets of  
the unlabelled sample space $\Omega$. For each covariant, measureable
set, $A$, the dynamics provides a number $\mu(A)$, the probability that the
realised causet is an element of $A$. But what is a physical, {\it i.e.}
label-independent characterisation of the elements of such a set $A$?  

We now know that given almost any 
causet $c$ in $\Omega$ we can determine whether or not it is in $A$ by 
asking countable logical combinations of questions of the form, ``Is the 
finite poset $b$ a {\it stem}\footnote{A stem of a causet is a finite subcauset
which contains all its own ancestors.} in $c$?'' The only case in which 
this is not possible is 
when $c$ is a so-called rogue causet for which there exists a 
non-isomorphic causet, $c'$ with exactly the same stems as $c$.  
The set of rogues, however, has measure zero and so they almost 
surely do not occur and we can simply eliminate them from the 
sample space.

Which finite posets are  
stems in $c$ is manifestly 
label-invariant data and the result implies that
asking these sorts of questions 
(countably combined using ``not'', ``and'' and ``or'') 
is sufficient to exhaust all that the dynamics can tell us. 

This result depends crucially 
on the dynamics. It would not hold for other sequential 
growth models in which the special type of causet which 
threatens to spoil the result does not have measure zero. So, though we 
can claim to have solved the ``problem of covariance'' for the Rideout-Sorkin
models, when we have a candidate quantum dynamics, we will have to 
check whether the ``stem questions'' are still the whole story. 

\subsection{The problem of Now}

Nothing  seems more fundamental to our experience of the 
world than that we have those
experiences ``in time.'' We feel strongly that the moment of Now 
is special but are also aware that it cannot be pinned down: it is elusive 
and constantly slips away as time flows on.   
As powerful as these feelings are, they seem to be contradicted
by our best current theory about time, General Relativity. There is 
no scientific contradiction, no prediction of General Relativity
that can be shown to be wrong. But, the general covariance
of the theory implies that the proper way to think of spacetime is 
 ``timelessly'' as a single entity, laid out once and for all
like the whole reel of a movie. There's no physical place for any 
``Now'' in spacetime and this seems at odds with our 
perceptions.

In the Rideout-Sorkin sequential growth models we see, 
if not a resolution, then at least an easing of this tension. 
The models are covariant, but nevertheless, the dynamics is 
specified as a sequential growth. An element is born, another 
one is born. There is growth and change. 
Things happen! But the general covariance means that the
physical order in which they happen is a {\it partial}
order, not a total order. This doesn't give any physical 
significance to a {\it universal} Now, but rather to 
events, to a Here-and-Now.

I am not claiming that this picture of accumulating events
(which will have to be reassessed in the quantum theory) 
would {\it explain} why we experience
time passing, but it is more compatible with our experience
than the Block Universe
view.

One might ask whether such a point of view, that events happen in a 
partial order, could be held  within General Relativity itself. 
Certain consequences have to be grappled with: if one event, $x$ has
occurred, then before another event, $y$ can occur to its future, infinitely 
many events must occur in between and this is true no matter 
how close in proper time $x$ and $y$ are. Perhaps it is 
possible to make this coherent but, to my mind, the discreteness 
of causal sets makes it easier 
to make sense of this picture of 
events occurring in a partial order. 

 Another thing left unexplained by General Relativity 
is the observational fact of the inexorable nature of time: it 
will not stop. Tomorrow will come. We have not so far encountered
an ``edge'' in spacetime where time simply comes to an end 
(nor, for that matter, an edge in space). Spacetimes with boundaries, 
for example a finite portion of Minkowski space, are 
just as much solutions of the Einstein equations as those without
and General Relativity cannot of itself account for the lack of
boundaries and holes in spacetime.  
In a Rideout-Sorkin universe, 
one can prove that time cannot stop. 
A sequential growth must be run to infinity if it is to 
be generally covariant and Joe Henson has 
proved that in a Rideout-Sorkin model every element has 
a descendant and therefore infinitely many descendants.\cite{Henson:2002}  
If such a result continues to hold true in the quantum case, not only 
would it prove that tomorrow will always come but also would imply that 
in situations where we expect there to be no continuum approximation 
at all, such as a Big Crunch or the singularity of a black hole, 
the causal set will continue to grow afterwards. 

The Rideout-Sorkin models 
give us a tantalising glimpse of the sorts of fundamental questions 
that may find their answers in quantum gravity when we have it. 

\subsection{The quantum case}

The mathematical structure of ordinary unitary Quantum Mechanics in its 
sum-over-histories formulation is 
a generalisation of a classical stochastic theory in
which instead of a probability measure there is 
a ``quantum measure'' on the sample space.\footnote{Indeed
Quantum Mechanics is the
first of an infinite series of generalisations resulting
in a heirarchy of measure theories with increasingly complex 
patterns of ``interference'' between histories.\cite{Sorkin:1994dt}}
 The quantum 
measure of a subset, $A$, of $\Omega$ is calculated in 
the familiar way by summing and then squaring the 
amplitudes of the fine grained elements of $A$. A quantum 
measure differs from a classical probability measure in 
the phenomenon of interference between histories which leads 
to it being non-additive. A familar 
example is the double slit experiment: the quantum measure of the
set of histories which go through the upper slit plus the 
quantum measure of the set of histories which go through the 
lower slit is not equal to the quantum measure of the set of 
all histories which end up on the screen. 

After the breakthrough of the Rideout-Sorkin models, it seemed
that it would be relatively straightforward to generalise the 
derivation to obtain a quantum measure for causets.
Roughly speaking, instead of 
transition probabilities there  would be transition amplitudes
and the quantum measure would be constructed from them 
via a generalisation of ``sum and square'' appropriate for a 
non-unitary dynamics.\cite{Martin:2004xi} 
General covariance and the appropriate quantum version of the Bell
Causality condition  could then be solved to find the form that the 
transition amplitudes must take. However, it is proving
difficult to find the required Quantum Bell Causality condition, not
least because the condition is not known even in the case of 
ordinary unitary Quantum Mechanics in a background with
a fixed causal structure though we do now have at least a candidate 
for it.\cite{Craig:2005}

Even if we had in hand a covariant, causal 
quantum measure, there would still remain the problem of interpreting 
it. The interference between histories and 
consequent non-additive nature of the quantum 
measure mean that we are exploring new territory here.  
Reference \cite{Sorkin:1995nj} is a first attempt at a realist 
interpretation for quantum measure theory. 
It relies on the adequacy of predictions such as, 
``Event $X$ is very very unlikely to occur,'' to cover 
all the predictions we want to make, which should 
include all the predictions we can make in standard quantum mechanics
with its Copenhagen 
Interpretation.  
This adequacy is 
explicitly denied by Kent\cite{Kent:1989nm} and I have tended to 
agree with this judgement. However, the quantum measure is the 
result of taking a conservative approach to Quantum Mechanics 
(no new dynamics) whilst making histories primary,  maintaining 
fundamental spacetime covariance and taking a completely realist 
perspective.    As such it deserves to be persevered with.

\section{Conclusions} 

The belabouring, in section 2, of the correspondence between 
the inverse processes of discretisation and continuum approximation
makes manifest a certain conservatism of causal set theory: 
the steps are familiar, we've been 
down similar roads before. Moreover, spacetime, albeit discrete,
is still considered to be real. There is no replacement of spacetime 
by a substance of a completely different ilk, 
such as a collection of $D0$-branes 
in 11-dimensional flat spacetime\cite{Banks:1996vh} to choose an 
example at random, as the 
underlying ontology. The radical kinematical input is 
the postulate of fundamental discreteness. 

However, no matter how smooth one can make such arguments for causal sets,
no scientific theory can be arrived at by pure philosophical
introspection. For a start, hard scientific labour is already 
contained in the proof of the essential result that the causal 
structure fixes the spacetime metric up to local volume information. 
This has been strengthened by results that show that
topological and  geometrical 
information can indeed be ``read off'' from a 
causet which is a sprinkling into Minkowski spacetime. 
More importantly, we need to do a great deal of further work.
For example, within kinematics
the Hauptvermutung needs to be given a formal mathematical 
statement and more evidence provided for it and we need to have
more results on how to read off geometrical information from a 
causet especially in the case of curved spacetime.

In the area of dynamics
the Rideout-Sorkin models, though only classically stochastic, 
are proving invaluable for exploring issues such as 
the problem of general covariance. It is possible that 
a more or less direct generalisation of the derivation of these
models will lead to the desired quantum theory. Finding a 
quantum dynamics is 
the central challenge for workers in causal set theory. I have given a
somewhat sketchy account of some of the conceptual hurdles that 
need to be overcome before this can be achieved.

One thing that has not been explored in this article is
how far causal set theory has come in the area of
{\it phenomenology}, in other words in the deriving of 
observable consequences of the theory. It will be hard for 
any  approach to quantum gravity to come to be universally accepted 
without experimental and observational evidence in 
the form of predictions made and verified. In this regard,
causal set theory already has the advantage of a long-standing 
prediction of the current order of magnitude of the 
cosmological constant, or  
``dark energy density''\cite{Sorkin:1990bh,Sorkin:1990bj,Sorkin:1997gi} 
that has apparently now been verified. 
The argument leading to this prediction is heuristic -- it depends on 
certain expectations about the quantum theory -- and can only
be made rigorous with the advent of the full quantum causal set dynamics.
However, the sheer unexpectedness of the observational result 
amongst the wider community of theorists
-- some cosmologists have called it preposterous\cite{Carroll:2001xs}
-- is great encouragement
to investigate the arguments further. Numerical simulations of 
stochastic cosmologies based on the arguments bear out the conclusion that 
the envelope of the fluctuations of the dark energy density
tends to ``track'' the energy density of matter \cite{Ahmed:2002mj}. 
Improvements would include models which allow spatial inhomogeneities 
in the dark energy density.

 Moreover, there is
great promise for further phenomenological model building.
The unambiguous kinematics of the causal set approach 
means that there is an obvious way to try to make phenomenological 
models: create a plausible dynamics for 
matter (particles or fields, say) on the background of
a causal set that is well approximated by the classical spacetime that we
actually observe: Minkowski spacetime or Friedmann-Robertson-Walker
spacetime, depending on the physical context. The dynamics of the
matter might be classical or quantum. The limitation of
such model building is that it doesn't take into account
the quantum dynamics of the causal set itself, nor any
back-reaction, but these models
could be a first step in deriving observable effects of the
underlying discreteness on phenomena such as the
propagation of matter over long  distances. An example of 
exactly this form is a model of point particle motion on a 
causet which leads to a prediction of a Lorentz invariant 
diffusion in particle momentum  and therefore energy.\cite{Dowker:2003hb} 
A naive application to cosmic  protons doesn't work as a universal 
acceleration mechanism that might explain high cosmic energy rays 
but a quantum version might do better and 
the idea could be applied to other particles like neutrinos. 

In causal set theory, we now have 
the mathematical structure 
that Einstein lacked, giving us a framework for a 
fundamentally discrete theory of spacetime which 
does not rely on any continuum concept as an aid. How successful 
it will be in realising the unification that Einstein 
hoped for, will be for the future to  decide. But let me
end by musing on a unification even beyond that of
quantum gravity: the unification of kinematics and
dynamics. In causal set theory as currently conceived,  
the subject of the theory and the laws by which it is governed 
are different in kind. This is apparent in the Rideout-Sorkin models
for example. The law of growth is given by a sequence of non-negative 
numbers. This law is not part of physical reality which is the causal set. 
To a materialist like myself, it would be more satisfying if the 
laws themselves were, somehow,
physically real; then the physical universe, meaning
everything that exists, would be 
``self-governing'' and not subject to laws imposed on it 
from outside.  
Should these nebulous ideas find concrete expression 
it would represent perhaps the ultimate 
unity of physics.

\section*{Acknowledgments}
\addcontentsline{toc}{section}{Acknowledgements}
I gratefully acknowledge the influence of
Rafael Sorkin on this article. Anyone familiar with his 
work will realise that I have 
plagiarised his writings shamelessly. I thank him and 
my collaborators on causal sets, Graham Brightwell, Raquel Garcia, 
Joe Henson, Isabelle Herbauts and Sumati Surya for
their insights and expertise. 

\bibliographystyle{wsabbrvnat}
\bibliography{../../../Bibliography/refs}

\end{document}